\documentstyle[11pt,newpasp,twoside,psfig]{article}
\def\edcomment#1{\iffalse\marginpar{\raggedright\sl#1\/}\else\relax\fi}
\reversemarginpar

\begin{document}
\title{{\sc HI} Survey Science with the Canadian Large Adaptive Reflector}
\author{S. C\^ot\'e}
\affil{Herzberg Institute of Astrophysics, NRC, 5071 West Saanich Rd, Victoria, B.C., V9E 2E7, Canada}
\author{A.R. Taylor}
\affil{Department of Physics and Astronomy, University of Calgary, Canada}
\author{P.E. Dewdney}
\affil{Herzberg Institute of Astrophysics, NRC,  PO Box 248, Penticton, B.C., 
V2A 6K3, Canada}

\begin{abstract}

The Canadian Large Adaptive Reflector (CLAR) is a proposed prototype of
a new concept for large, filled-aperture radio telescopes. 
The prototype would have a 300-metre aperture, working up to frequencies
of at least 1.4 GHz, and would be equipped with a multi-beam phased array
providing a field-of-view of 0.8$^{\circ}$ at that frequency.
The largest fully-steerable radio telescope in the world,
and endowed with a large field-of-view, the CLAR will be uniquely suited for
deep spectral imaging over large areas of the sky. 
Conducted over a period of three to four years, a CLAR Northern-Sky
Survey would allow us to simultaneously: 
survey at arcminute scales the distribution and kinematics of the
faint HI in the halo of the Milky Way and High Velocity
Clouds;
chart the large scale distribution of galaxies in HI out to
redshift close to 1;
reveal the structure and dynamics of the cosmic web responsible
for wide-spread Lyman $\alpha$ absorption systems;
image the signal of the reionization of the Universe over a
large area with resolution of 10's of arcminutes.

\end{abstract}

\section{Introduction}

The Large Adaptive Reflector (LAR) is a technical concept for a very
large radio telescope that is under development in Canada for 
potential use for the Square Kilometre Array.
A complete technical description of the LAR can be found in Dewdney 
(2000) and Legg (1998).  
A technical concept for a proposed prototype, the Canadian Large
Adaptive Reflector, is under development.
The CLAR would consist of a 300-metre parabolic reflector, 
with a feed, 
suspended by a balloon and controled by tethers, consisting of a phased
array with 2300 elements synthesising into 150 primary beams,  
providing a field-of-view of 0.8$^{\circ}$ with a resolution of 2 arcmin 
at a frequency of 1.4 GHz. 
The largest fully-steerable radio telescope in the world,
and endowed with a large field-of-view, the CLAR will be uniquely suited for
deep spectral imaging over large areas of the sky. 
Figure 1 shows the mapping efficiency of existing and planned radio 
telescopes as a function of angular resolution, $\theta$. The
mapping speed is presented as solid angle imaged to a sensitivity of
1 K in a 0.8 km/s channel in one hour.  The plot was adapted from
a plot originally created by John Dickey (private communication).
While arrays, because of their distributed aperture, have a mapping 
speed dropping off as $\theta ^4$, filled-aperture telescopes follow 
$\theta ^2$ due to their higher brightness sensitivity.
Multi-beam feeds arrays such as that at Parkes, and with CLAR,
can provide a large field of view, despite the smaller beams (higher
resolution) of larger aperture single-dish antennas.
The CLAR will therefore be the fastest spectropic survey radio telescope
while at the same time providing angular resolution of a
few arcminutes at 1420 MHz.

\begin{figure}
\centerline{\vbox{
\psfig{figure=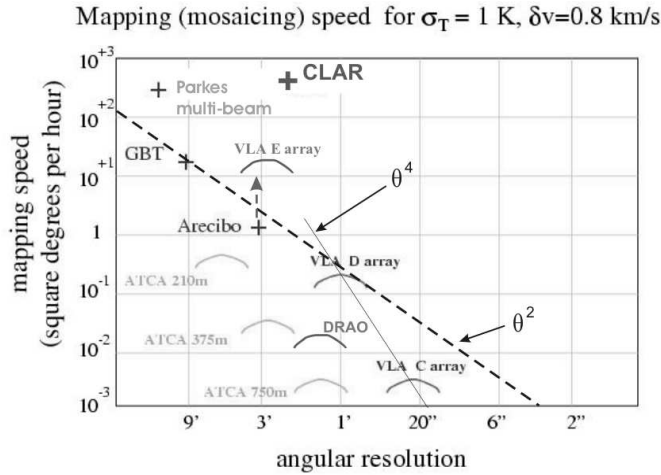,width=8.8cm}
}}
\caption{The mapping efficiency of current and future radio telescopes
versus angular resolution. }
\end{figure}

\section{CLAR Surveys}

The CLAR would be operated as a wide-area survey telescope.
Two distinct Surveys are planned:

\begin{itemize}

\item{} An All-Sky Northern Hemisphere Survey, conducted over a period of 
about 3 years, with 
 an effective integration of 2300s per pointing (corresponding to a 3$\sigma $
HI mass detection limit of about 5$\times 10^{10}$M$_\odot$ to z=1)

\item{} A Deep Field Survey, over a 50 square degrees region, requiring two years, with 100 hours per 
pointing (down to roughly 5$\times 10^{9}$M$_\odot$ to z=1)

\end{itemize}

Each of these surveys will be carried out 
with several back-ends at the same time, to acquire data simultaneously in
different modes:

\begin{itemize}

\item{} 10000 spectral line channels over 700 MHz to 1400 MHz providing HI imaging from
z=0 to 1 with sensitivity of 2 mK at a velocity resolution of 13 km/s.

\item{} 1024 spectral line channels at z=0 to provide a velocity resolution of 1 km/s
to map HI in the Galaxy and the Local Universe

\item{} another feed and receiver to sample some continuum bands at $<$700 MHz

\item{} a pulsar back-end.

\end{itemize}

\section{Science Goals}

We will concentrate in this section mainly on the HI science that these 
surveys will
enable us to explore. The CLAR All-sky Northern Hemisphere survey will be more
than 10 times more sensitive than what is presently underway (see details on the
Jodrel Bank HIJASS survey, this volume),  as well as offering better 
resolutions 
both spatially (2' instead of 12') and in velocity (1 km/s for the Local 
survey, compared
to 18 km/s).

\subsection{Probing the Galactic Halo}

The Survey will trace in details the distribution and kinematics of the
faint HI layer between the disk and halo of the Milky Way, revealing the 
physics of mass and energy transfer between the disk and halo, as well as 
map in details the High Velocity
Clouds, and the HI in Local Group objects.
At the same time the multiple continuum bands between 700 and 1400 MHz
will provide an image of the total power of the continuum emission and the
spatial variation of the spectral index, with a receiver allowing for 
extremely accurate relative photometric calibration. This would be invaluable
for deconvolving the foreground emission from upcoming space missions to 
measure the arcminute scale structure of the Cosmic Microwave Background
radiation (PLANCK).
 
\subsection{Defining the HI Large Scale Structure}

 The surveys will chart the large scale distribution of galaxies in HI out to
redshift close to 1, complementing and extending the Sloan Survey at
optical wavelengths. All galaxy types are detected in HI, 
and the survey will be able to detect HI masses as low as 
$10^3$M$_{\odot}$ D$^2_{Mpc}$.
Moreover the CLAR surveys will provide redshifts for galaxies biased against or 
even completely missed by optical surveys, like Low-Surface-Brightness galaxies
(LSBs) and dwarf galaxies. For example the Sloan survey will get follow-up 
spectroscopy only on objects of surface brightness 24 mag/arsec$^2$ or 
brigther.
Below $10^7$M$_{\odot}$ the HI mass function is very ill-defined,
the slope needs to be better constrained as there might be a much larger 
number of low mass galaxies. For example in the nearby Sculptor Group C\^ot\'e 
et al (1997) found that 75\% of the dwarf galaxies detected in HI are below
2$\times 10^7$M$_{\odot}$. 

Using the HI linewidths, this will also provide us with an estimate of the
maximum rotation velocities of the galaxies and hence their mass, 
therefore tracing the dark matter content for all the disk galaxies in the
sample. This will enable us to construct a galaxy mass function, of great 
interest to compare with cosmological scenario predictions.

The survey will also trace the HI gas between galaxies. HI arms and tidal 
features will be mapped in groups and clusters, tracing out the history
of gravitational interactions between galaxies. 
 
\subsection{Detecting the Cosmic Web}

Perhaps most interestingly the CLAR Surveys column density limit 
will be low enough to
study very low column density gas in the Universe never before detected directly
in emission. Spectra of QSOs reveal large numbers of 
Lyman $\alpha$ absorption lines, due to absorption by foreground condensations 
of gas along their line-of-sight. 
It is not known if this absorbing gas is related to huge diffuse HI halos 
of galaxies, 
or HI haloes of undetected LSBs or dwarf galaxies, or intragroup gas. On the other hand 
hydrodynamical simulations  predict the existence of a Cosmic Web, a 
complicated 
network of gaseous filaments and sheets, interconnecting overdense regions 
where galaxies formed in the underlying dark matter potential.  
The CLAR All-Sky survey will detect in emission HI gas with column densities
down to 10$^{17}$cm$^{-2}$, while the deep survey will reach 10$^{16}$cm$^{-2}$.
This is well within the range of HI column densities of many measured 
Ly$\alpha$ lines. With the CLAR the structure of the gas giving rise to these
Ly$\alpha$ lines and its connection to galaxies will be revealed for the 
first time.
 
\subsection{Mapping the Reionization Epoch}

If the reionization of the Universe occured at 5$< z <$20 then it should be 
detectable as a `step' in the radio spectrum between 70 and 240 MHz due to
redshifted HI emission. The CLAR will have the sensitivity to detect 
this step and, more interestingly, to map the structure of the signal on scales
of 10's of arcminutes, revealing the causes of the reionization. For example if
it is caused by a large number of stars well distributed in space the step will
be abrupt over the whole sky (like a phase transition of the whole universe).
But if it is caused by rare quasars it will appear irregular spatially as 
well as 
appearing at different redshifts in different directions.
The peak intensities of the signal should be readily detectable by the 
CLAR (see  
Tozzi et al 2000; Tozzi, this volume).

\section{Conclusions}

The CLAR HI Surveys will give us our clearest view of the HI Universe.
It will map the large scale structure of galaxies, trace the cosmic web,
and map the reionization epoch. The results of these Surveys would be
central to the planning of the future SKA Science surveys.

\end{document}